\documentclass[epj]{svjour}
\usepackage{graphics,psfig}

\newcommand{\gtrsim}{
\,\raisebox{0.35ex}{$>$} \hspace{-1.7ex}\raisebox{-0.65ex}{$\sim$}\, }

\newcommand{\lesssim}{
\,\raisebox{0.35ex}{$<$} \hspace{-1.7ex}\raisebox{-0.65ex}{$\sim$}\, }
\begin{document}
\title{Tunneling and EPR linewidths due to dislocations in Mn$_{12}$ acetate}
%
\author{D. A. Garanin  \inst{1,2} \and E. M. Chudnovsky \inst{1}
}                     
%
%
\institute{Department of Physics and Astronomy, Lehman College, City University of New York, \\
250 Bedford Park Boulevard West, Bronx, New York 10468-1589 \and Institut f\"ur Physik,
Universit\"at Mainz, D-55099 Mainz, Germany}
\date{Received: 15 March 2002}
%
\abstract{
We compute the width and shape of the EPR and tunneling resonances due to dislocations in
Mn-12 acetate crystals. Uncorrelated dislocations produce the Gaussian shape of resonances
while dislocations bound in pairs produce the Lorentzian shape. We stress that the uniaxial
spin Hamiltonian together with crystal defects can explain the totality of experimental data
on Mn-12.
}
\maketitle

The discovery of resonant spin tunneling in Mn$_{12}$ acetate
\cite{frisartejzio96,tejadabarbara} has triggered an avalanch of theoretical and experimental
works on molecular nanomagnets (see Ref.\ \cite{garchu02prb} for references).
Despite of a significant progress made in understanding Mn$_{12}$ and later
discovered Fe$_{8}$ spin-10 systems, a number of key questions remains
unanswered.
One of them is the width and shape of the tunneling resonance.
In the past, attempts were made to explain them by phonons \cite{garchu97,leulos9900}, nuclear
spins \cite{prosta98}, and dipolar fields \cite{prosta98,ohmsanpau98,cucetal99,alofer01}.
Recently, we have suggested that quantum magnetic relaxation in molecular nanomagnets can be
explained by dislocations in the crystal lattice \cite{chugar01,garchu02prb}.
Four recent experimental works give evidence of the effect of defects on
tunneling and EPR in single crystals of Mn$_{12}$ and Fe$_{8}$
\cite{parksetal01,parketal01,mertesetal01,hertortejmol01}, in accordance with
our suggestions.
The analysis of these experiments requires computation of the width and shape
of EPR and tunneling resonances due to dislocations, which is done in this
Letter.
We argue that dislocations at common concentrations provide the observed width
of tunneling resonances and the observed width of the EPR in Mn$_{12}$ and
Fe$_{8}$.

Qualitatively, the importance of dislocations is clear from the fact that they
give rise to long-ranged elastic strains which modulate crystal fields and thus
create spatial dependence of the magnetic anisotropy.
In spin tunneling and EPR experiments the resonant values of the magnetic field
are determined by the anisotropy constants.
For Mn$_{12}$ crystals in the field parallel to the easy axis, with the
Hamiltonian ${\cal H} = -DS_z^2-H_zS_z + {\cal H}'$ (${\cal H}'$ being a small
tunneling term), the resonant spin tunneling occurs at
\begin{equation}\label{ResCond}
H_z=kD, \qquad k=0,\pm 1, \pm 2,\ldots, \pm(2S-1),
\end{equation}
while the EPR between the levels $m$ and $m-1$ at frequency $\omega$ occurs at
\begin{equation}\label{EPRCond}
H_z = \omega - D(2m-1).
\end{equation}
In our two recent works \cite{chugar01,garchu02prb} we computed ${\cal H}'$ due to
dislocations and neglected the effect of dislocations on the resonance condition.
Meantime, the spatial dependence of the magnetic anisotropy $D$ due to
dislocations causes resonances to spread over a certain field range.
This range depends on the magnetoelastic coupling, the type, and concentration
of dislocations.

The terms in the magnetoelastic coupling that are responsible for the
modulation of the uniaxial anisotropy constant $D$ can be written as
\begin{equation}\label{Hme}
 {\cal H}_{\rm me} = -{D'} S_z^2,
\qquad {D'} = D[g_0 ( \varepsilon_{xx} + \varepsilon_{yy} ) - g'_0 \varepsilon_{zz} ],
\end{equation}
where
\begin{equation}\label{epsDef}
\varepsilon_{{\alpha}{\beta}}=\frac{1}{2}\left(\frac{\partial u_\alpha}
{\partial x_ \beta} + \frac{\partial u_ \beta }{\partial x_\alpha} \right)
\end{equation}
is the linear deformation tensor and $\alpha,\beta = x,y,z$.
The coupling constants $g_0$ and  $g'_0$ must be of order one, see Ref.\
\cite{leulos00epl} and references therein.
For illustrations, we will use  $g_0=g'_0=1$.

\begin{figure}[t]
\unitlength1cm
\begin{picture}(7,5)
\psfig{file=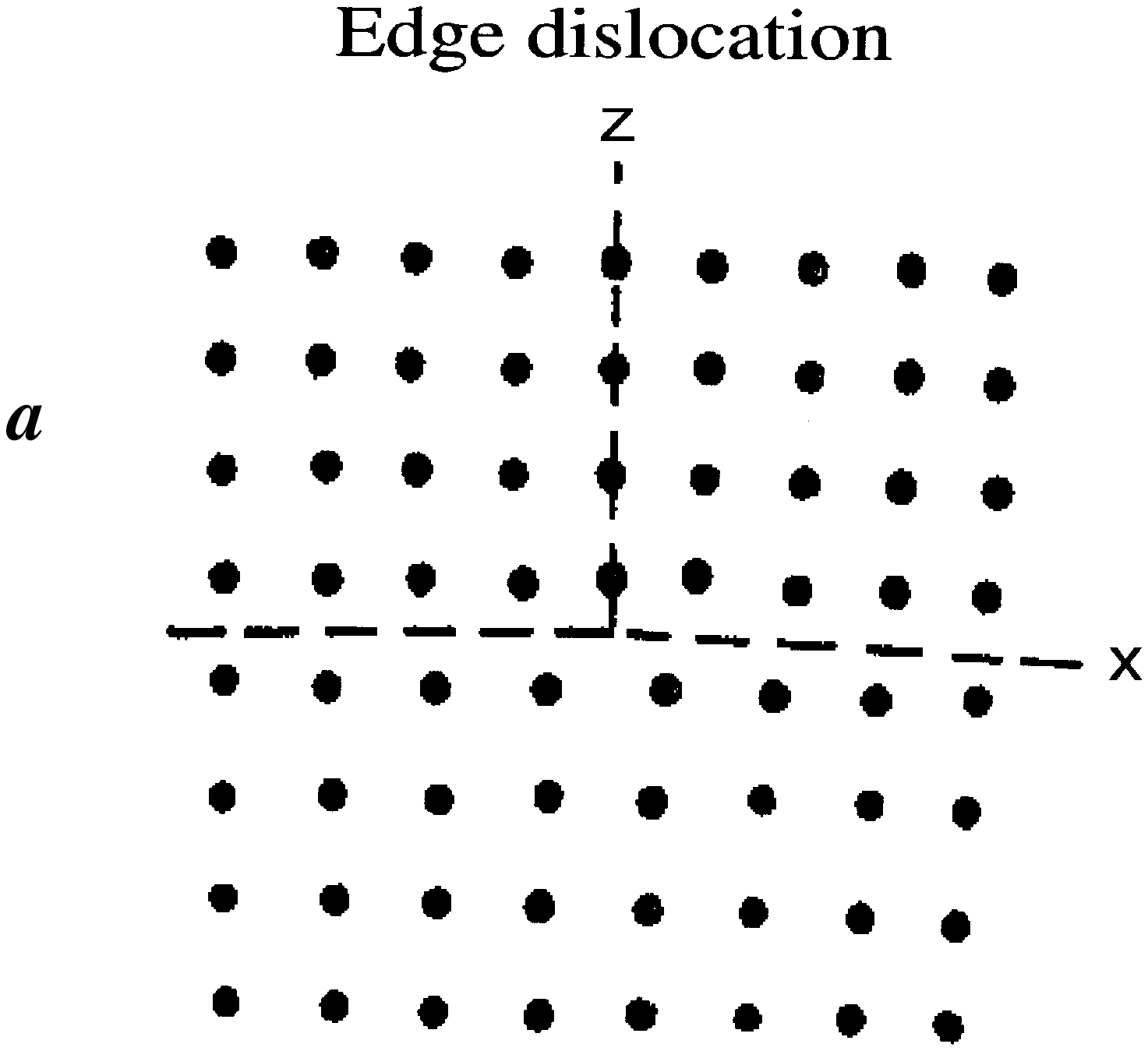,angle=-90,width=9cm}
%
\end{picture}
\begin{picture}(7,5)
\psfig{file=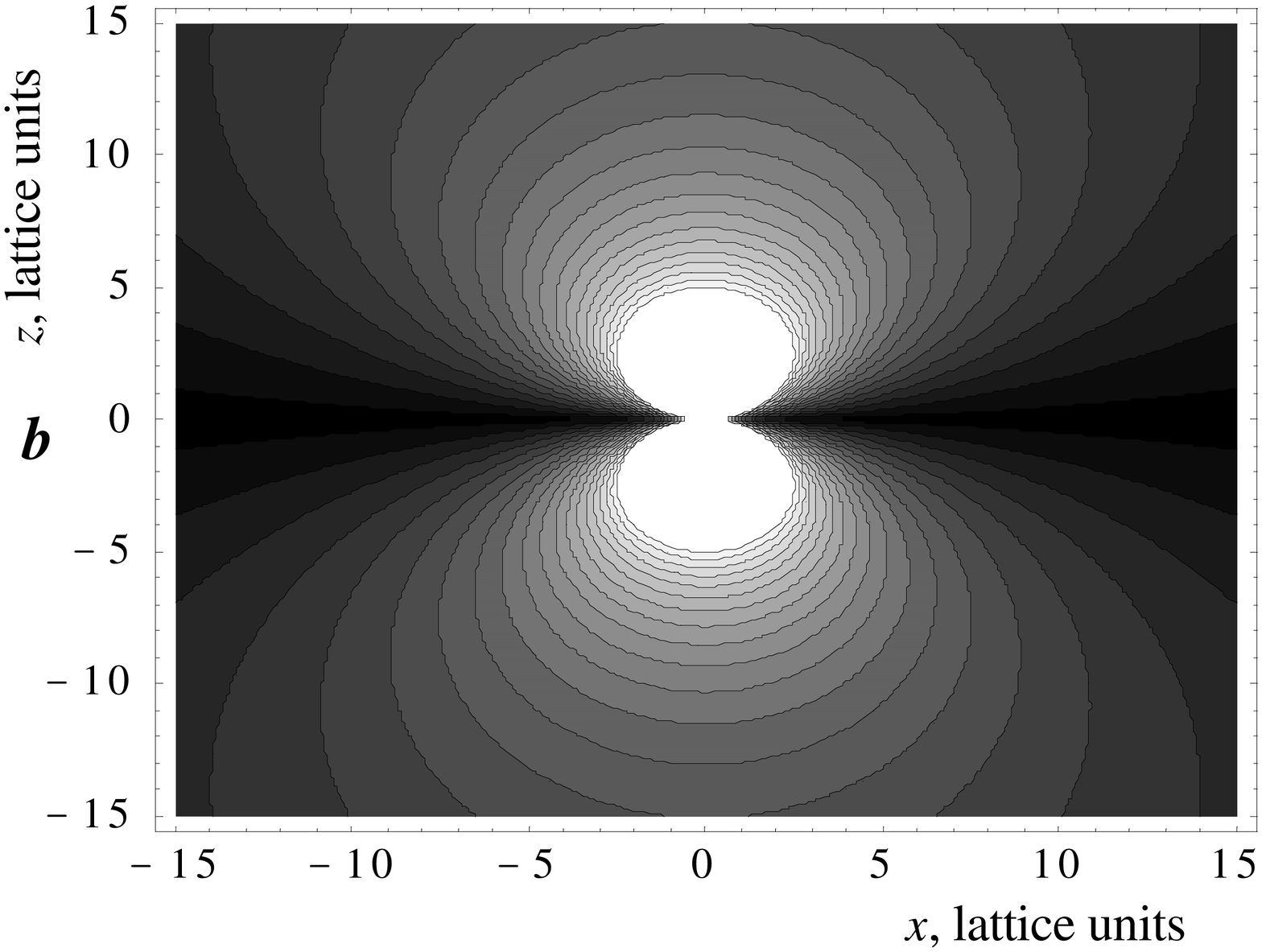,angle=-90,width=7.5cm}
\end{picture}
\caption{ \label{fig_edge}
($a$): Edge dislocation running along the $y$ axis with the extra plane $y,z$
inserted at $z>0$; ($b$): The magnitude of the dislocation-induced
contribution into the uniaxial anisotropy constant $D$. }
\end{figure}

For screw dislocations, one has $\varepsilon_{xx} = \varepsilon_{yy} =
\varepsilon_{zz}=0$ so this type of dislocations does not contribute into the
EPR and tunneling resonance linewidths.
For edge dislocations with the axis along the $z$ axis of the crystal and the extra plane
$z,y$ inserted at $y>0$ (see Fig.\ \ref{fig_edge}) one has $u_z=0$, whereas other displacement
components are given by  \cite{lanlif7}
\begin{equation}\label{xDisplEdgeZ}
u_x = \frac b {2\pi} \left[ \arctan\frac y x + \frac 1 {2(1-\sigma)} \frac {xy}
{x^2 + y^2} \right]
\end{equation}
and
\begin{equation}\label{yDisplEdgeZ}
u_y = -\frac b {2\pi} \left[ \frac{1-2\sigma}{4(1-\sigma)} \ln(x^2 + y^2) +
\frac 1 {2(1-\sigma)} \frac {x^2} {x^2 + y^2} \right],
\end{equation}
where $b$ is the Burgers vector coinciding with the lattice period and
$0<\sigma <1/2$ is the Poisson elastic coefficient (we will use $\sigma=0.25$
in the numerical work).
The relevant components of the deformation tensor are $\varepsilon_{zz}=0$ and
\begin{equation}\label{epsxxyyEdgeZ}
\varepsilon_{xx} + \varepsilon_{yy} = -\frac{b(1-2\sigma)}{2\pi (1-\sigma)}
\frac y {x^2+y^2}.
\end{equation}

Displacements due to other types of edge dislocations can be obtained from Eqs.\
(\ref{xDisplEdgeZ}) and (\ref{yDisplEdgeZ}) by the change of variables.
In particular, for edge dislocations with the axis along the $y$ axis of the
crystal and the extra plane $z,y$ inserted at $z>0$ one should make a
replacement $y\Rightarrow z$.
This yields $\varepsilon_{yy}=0$,
\begin{equation}\label{epsxxEdgeY}
\varepsilon_{xx}  = -\frac{b}{4\pi (1-\sigma)}  z \frac
{(3-2\sigma)x^2+(1-2\sigma)z^2} {(x^2+z^2)^2},
\end{equation}
and
\begin{equation}\label{epszzEdgeY}
\varepsilon_{zz}  = -\frac{b}{4\pi (1-\sigma)}  z \frac
{(1+2\sigma)x^2-(1-2\sigma)z^2} {(x^2+z^2)^2}.
\end{equation}

Generally, the axis of an edge dislocation can be directed along the $x$, $y$,
and $z$ axes of the crystal, and in each of these cases there are four possible
orientations of the extra crystallographic plane.
One can write
\begin{equation}\label{DGen}
D' = D\frac {g_D(\varphi)}{r}.
\end{equation}
Here $r$ is the distance from the dislocation axis, measured in the lattice
units, whereas $g_D(\varphi)$ is a function of the angle which is of order one
if $g_0 \sim g'_0 \sim 1$.
One can immediately see from Eq.\ (\ref{DGen}) that the contribution of dislocations in the
EPR and tunneling resonance linewidth must be large.
Indeed, for $r\sim 1$ one has $D'\sim D$, whereas the spatial decay of $D'$ is
slow, so that each dislocation affects a large number of molecules in the
crystal thus rendering each molecule a different value of the uniaxial
anisotropy.
This leads to a substantial inhomogeneous broadening of resonances which follows from Eqs.\
(\ref{ResCond}) and (\ref{EPRCond}).

In a crystal with dislocations, the deformation tensor at any given point is a
sum of contributions due to many different dislocations.
The superposition principle for deformations follows from the linearity of the
equations of the theory of elasticity \cite{lanlif7} and it holds everywhere
outside dislocation cores, {\it i.e.}, for the distances from the dislocation
axes $r \gtrsim 1$.
Statistical properties of deformations in a crystal depend on the spatial
distribution of dislocations which is poorly known.
Let us find analytically the distribution of the anisotropy constant $D$
assuming that dislocations are distributed at random.
The distribution function for $D'$ in Eq.\ (\ref{Hme}) can be defined as
\begin{equation}\label{fDDef}
f_{\tilde D'} = \left\langle \delta \left(\tilde D' - \sum_{i=1}^N \tilde
D'({\bf r}-{\bf r}_i) \right) \right\rangle, \qquad \tilde D' \equiv \frac {D'
} D,
\end{equation}
where $N \gg 1$ is the number of dislocations in the crystal and the averaging
is carried out over their positions ${\bf r}_i$ in the plane perpendicular to
the dislocation axis within a circular region of radius $R$.
We choose the observation point in the middle of the crystal, ${\bf r}=0$.
One can define
\begin{equation}\label{RcDef}
c = \frac N {\pi R^2} = \frac 1 {\pi R_c^2},
\end{equation}
where $c$ is the concentration of dislocations and $R_c$ is the characteristic
distance between dislocations.

Let us at first analyze the large-$|\tilde D'|$ asymptotes of $f_{\tilde D'}$
due to the regions with large deformations of both signs close to one of
dislocations.
In that case one can neglect the influence of all other dislocations and
consider the one-dislocation model
\begin{equation}\label{fDoneDef}
f_{\tilde D'} =  \frac 1 {\pi R_c^2} \int_0^{2\pi} d\varphi \int_0^{R_c} rdr
\delta \left(\tilde D' - \frac{g_D(\varphi)} r \right).
\end{equation}
Integration yields
\begin{equation}\label{fDoneRes}
f_{\tilde D'} = \frac{ (\tilde D'_c)^2} { |\tilde D'|^3 }, \qquad |\tilde D'|
\gtrsim \tilde D'_c \equiv \frac{ \sqrt{\langle g_D(\varphi)^2\rangle}}{R_c},
\end{equation}
where $D'_c$ is the characteristic value of $D'$ at the distance $R_c$ and
$\langle\ldots\rangle$ is the angular average.
This formula becomes invalid for $\tilde D' \lesssim \tilde D'_c$, where the lines of constant
$\tilde D'$ in Eq.\ (\ref{fDoneDef}) cross the boundary of the region under consideration,
$r=R_c$.
In fact, for $\tilde D' \lesssim \tilde D'_c$  Eq.\ (\ref{fDoneDef}) becomes invalid and one
has to take into account other dislocations.
Eq.\ (\ref{fDoneRes}) suggests that one should introduce the distribution
function for the reduced quantity $\alpha$
\begin{equation}\label{falphaDef}
f_\alpha \equiv \tilde D'_c f_{\tilde D'}, \qquad \alpha \equiv\tilde D'/\tilde
D'_c,
\end{equation}
which has the asymptote
\begin{equation}\label{falphaAsymp}
f_\alpha = 1/|\alpha|^3, \qquad \alpha \gtrsim 1.
\end{equation}

In the general case, with the help of the identity $2\pi\delta(x) =
\int_{-\infty}^{\infty}d\omega e^{i\omega x}$, the averaging over the coordinates of different
dislocations in Eq.\ (\ref{fDDef}) can be factorized,
\begin{equation}\label{fDFact}
f_{\tilde D'} = \int_{-\infty}^{\infty} \frac{d\omega}{2\pi} e^{i\omega \tilde
D'} f(\omega)^N,
\end{equation}
where
\begin{equation}\label{fomegaDef}
f(\omega) \equiv  \frac 1 {\pi R^2} \int_0^{2\pi} d\varphi \int_0^{R} rdr
\exp\left( -\frac{i\omega g_D(\varphi)} r \right).
\end{equation}
In Eq.\ (\ref{fDDef}) we assumed for simplicity that all dislocations are of the same type.

\begin{figure}[t]
\unitlength1cm
\begin{picture}(11,6.5)
\centerline{\psfig{file=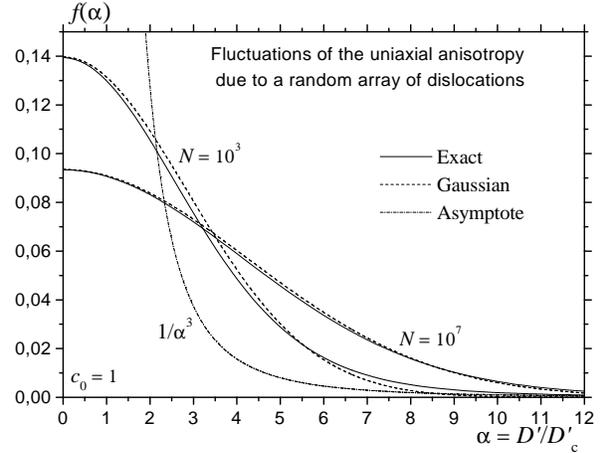,angle=-90,width=9cm}}
\end{picture}
\caption{ \label{fig_distribd} Distribution of the uniaxial anisotropy due to a
random array of edge dislocations}
\end{figure}

As we shall see, in Eqs.\ (\ref{fDFact}) and (\ref{fomegaDef}), $\omega \sim R_c \ll R$ for
$N\gg 1$, thus the argument of the exponential in Eq.\ (\ref{fomegaDef}) is small and
$f(\omega)$ is close to unity.
Then the exponential can be expanded and integrated, with a log accuracy, in
the interval $|\omega| \lesssim r < R$.
Given that $\langle g_D(\varphi)\rangle=0$, the result has the form
\begin{equation}\label{fomega}
f(\omega) \cong  1 - \frac{ \omega^2 \langle g_D(\varphi)^2\rangle }{R^2}
\ln\frac{c_0 R}{|\omega| \sqrt{\langle g_D(\varphi)^2\rangle}},
\end{equation}
where $c_0$ is a constant of order unity.
Now with the use of Eqs.\ (\ref{RcDef}) and (\ref{fDoneRes}) one can write
\begin{eqnarray}\label{fomegaN}
&&
 f(\omega)^N \cong 1 - \frac{ \omega^2 \langle g_D(\varphi)^2\rangle }{R_c^2} \ln\frac{c_0
R}{|\omega| \sqrt{\langle g_D(\varphi)^2\rangle}} \nonumber\\
 &&
 \qquad\qquad {}\cong
\exp\left[-(\omega \tilde D'_c)^2 \ln\frac{c_0\sqrt{N}}{|\omega| \tilde
D'_c}\right].
\end{eqnarray}
At this point one may forget about the initial assumption on the circular form
of the spatial region.
The shape of the crystal only affects the value of the constant $c_0$ under the
logarithm.
Eq.\ (\ref{fomegaN}) confirms the assumption $\omega \sim 1/\tilde D'_c \sim R_c$ made above.
Now we are prepared to write down the final result which is convenient to formulate in terms
of the function $f_\alpha$ defined by Eq.\ (\ref{falphaDef})
\begin{equation}\label{falphaRes}
f_\alpha \cong \frac 1 \pi \int_0^\Lambda du \cos(\alpha u) \exp\left( -u^2 \ln
\frac {c_0 \sqrt{N}}u \right).
\end{equation}
Here the cutoff $\Lambda$ satisfies $1 \ll \Lambda \ll \sqrt{N}$; one cannot
integrate up to $\infty$ since the form if the integrand is only valid for $u
\ll \sqrt{N}$.
Clearly, for large enough crystals with $N\gg 1$ the result does not depend on
$\Lambda$.
We remind that for the edge dislocations along the $Y$-axis, the distribution
of transverse anisotropies is an even function.
The distribution is shown for $\tilde D' > 0$ in Fig.\ \ref{fig_distribd}.

Integrating Eq.\ (\ref{falphaRes}) by parts three times, one can recover the asymptote of
$f_\alpha$ at $|\alpha|\gg 1$ which is given by Eq.\ (\ref{falphaAsymp}).
This power-law asymptote is a consequence of the logarithmic singularity of the integrand in
Eq.\ (\ref{falphaRes}) at $u\to 0$ and it leads to the divergence of the second moment of
$f_\alpha$.
On the other hand, for large $N$ the distribution function may be well
approximated by Gaussian for not too large $\alpha$.
Indeed, for large $N$ the logarithm in Eq.\ (\ref{falphaRes}) is weakly dependent on $u$ and
can be replaced by a constant.
The best value of this constant corresponds to $u$ for which the argument of
the exponential equals one.
This requires solving a transcendental equation that can be done in a
perturbative way.
With a good accuracy one can use
\begin{equation}\label{LogValue}
\ln \frac {c_0 \sqrt{N}}u\Rightarrow L =
\ln\left[c_0\sqrt{N\ln(c_0\sqrt{N})}\right]
\end{equation}
which results in the aproximation
\begin{equation}\label{Gaussian}
f_\alpha \cong \frac 1 {2\sqrt{\pi L}} \exp\left( - \frac {\alpha^2}{4L}
\right)
\end{equation}
which is also shown in Fig.\ \ref{fig_distribd}.

As the number $N$ of dislocations in the crystal inreases, the function $f_\alpha$ of Eq.\
(\ref{falphaRes}) becomes closer and closer to the Gaussian, whereas the power-law asymptote
given by Eq.\ (\ref{falphaAsymp}) becomes shifted to the region of very large $\alpha$ where
it is hardly visible.
This effect is due to the accumulation of small contributions from dislocations
situated at large distances from the observation point (of order of the linear
dimension of the crystal).
Such small contributions from distant dislocations, which lead to the Gaussian distribution
$f_\alpha$, win over contributions from close dislocations responsible for Eq.\
(\ref{falphaAsymp}).
Gaussian approximation for the function $f_{\tilde D'}$ with the help of Eq.\
(\ref{falphaDef}) can be written in the form
\begin{equation}\label{GaussianD}
f_{\tilde D'} \cong \frac 1 {2\tilde D'_{\tilde c} \sqrt{\pi}} \exp\left( -
\frac {\tilde D'^2}{(2\tilde D'_{\tilde c})^2} \right),
\end{equation}
where
\begin{equation}\label{DctilDef}
\tilde D'_{\tilde c} \equiv \tilde D'_c \sqrt{L} = \sqrt{\pi \langle
g(\varphi)^2\rangle \tilde c}, \qquad \tilde c \equiv cL.
\end{equation}
The standard deviation of $\tilde D'$ according to Eq.\ (\ref{GaussianD}) is $\sigma_{\tilde
D'}=\sqrt{2}\tilde D'_{\tilde c}$.
One can see that the accumulation of contributions from distant dislocations leads to the
effective logarithmic renormalization of the concentration of dislocations $c$ with $L$
defined by Eq.\ (\ref{LogValue}).
For edge dislocations running along the $z$-axis, the quantity $\sqrt{ \langle
g_D(\varphi)^2\rangle}$ is given by
\begin{equation}\label{gDAvrEdgeZ}
\sqrt{ \langle g_D(\varphi)^2\rangle} = \frac{g_0} {2\sqrt{2}\pi}
\frac{1-2\sigma}{1-\sigma},
\end{equation}
where $\sigma$ is the Poisson elastic coefficient.
For for $g_0=1$ and $\sigma=0.25$ one has $\sqrt{ \langle
g_D(\varphi)^2\rangle}\approx 0.075$.
For edge dislocations running perpendicular to the $z$-axis one obtains
\begin{eqnarray}\label{gDAvrEdgeY}
&&
 \sqrt{ \langle g_D(\varphi)^2\rangle} = \frac{1} {8\pi(1-\sigma)}
 \left[
8\sigma^2(g_0-g'_0)^2 \right.
\nonumber\\
&& \left. {}- 4\sigma(g_0-g'_0)(3g_0-g'_0) + 5g_0^2-2g_0g'_0+(g'_0)^2 \right]^{1/2},
\end{eqnarray}
which for $g_0=g'_0$ simplifies to $\sqrt{ \langle g_D(\varphi)^2\rangle} =
g_0/[4\pi(1-\sigma)]$.
For $g_0=1$ and $\sigma=0.25$ one has $\sqrt{ \langle
g_D(\varphi)^2\rangle}\approx 0.106$.

The experimentally studied Mn$_{12}$ crystals are rather large, about $0.5
\times 0.5$ mm$^2$, which corresponds to the cross-section of about $10^{11}$
lattice cells.
Even for the concentration of dislocations as small as $c=10^{-4}$ per cell,
the number of dislocation in the crystal is about $N\approx 10^7$.
For $c_0=1$ this gives $L=9.1$, {\it i.e.}, the effective concentration of
dislocations increases by an order of magnitude, $\tilde c = 0.91\times
10^{-3}$.
The corresponding value of $\tilde D'_{\tilde c}$ that follows from Eqs.\ (\ref{DctilDef}) and
(\ref{gDAvrEdgeY}) for the edge dislocations running perpendicular to the $z$ axis is $\tilde
D'_{\tilde c}= 0.567\times 10^{-2}$.
For $c=10^{-3}$ one obtains $L=10.3$, thus $\tilde D'_{\tilde c}= 1.91\times
10^{-2}$.
The renormalization of the concentration of dislocations and the Gaussian distribution of
transverse anisotropies for large crystals are clearly seen in Fig.\ \ref{fig_distribd}: The
distribution broadens in the $\alpha$-scale due to the increase $L$ with $N$, Eq.\
(\ref{LogValue}).

The dislocation mechanism proposed in this Letter can qualitatively explain the
experimentally observed tunneling \cite{frisarzio98} and EPR
\cite{parksetal01,parketal01} linewidths in Mn$_{12}$ Ac.
For the realistic concentrations of dislocations $c=10^{-3}$ the standard
deviation $\sigma_{\tilde D'}=\sqrt{2}\tilde D'_{\tilde c} \approx 0.027$ is in
accord with the fit $\sigma_D = 0.02 D$ of Ref.\ \cite{parketal01}.

Now we consider another model of distribution of dislocations in the crystal:
Dislocations of opposite signs bound into pairs at the distance $d$.
Distributions of this kind are more likely than a completely random
distribution since here the energy of elastic strains is lower.
At the distances $r\gg d$ from Eq.\ (\ref{DGen}) one obtains
\begin{equation}\label{DDGen}
D' = -D \frac d {r^2}  A(\varphi), \qquad A(\varphi)\equiv \frac
\partial{\partial\varphi} [ g_D(\varphi)\sin \varphi],
\end{equation}
where $\varphi$ is the angle between the vectors $\bf r$ and $\bf d$.
The calculations following after Eq.\ (\ref{fDDef}) should be now redone with $D'$ given by
Eq.\ (\ref{DDGen}) and the parameters $N$, $c$, and $R_c$ designating the number,
concentration, and the average distance between the {\em dislocation pairs}.
For $d\ll R_c \ll R$ the characteristic values of $\omega$ and $r$ are $\omega
\sim R_c^2/l$ and $r \sim \sqrt{\omega l} \sim R_c$.
The function $f(\omega)$ in Eq.\ (\ref{fDFact}) then reads
\begin{eqnarray}\label{fomegaDD}
&&
 f(\omega) \cong  1 - \frac 1 {\pi R^2} \int_0^{2\pi} d\varphi \int_0^{\infty} rdr \left[1 -
\cos \left(  \frac {\omega d} {r^2}  A(\varphi)\right)\right] \nonumber\\
&& \qquad
 {} = 1 - \frac {\pi l |\omega|}{2R^2} \langle |A(\varphi)|\rangle .
\end{eqnarray}
This results in
\begin{equation}\label{fomegaDDN}
f(\omega)^N \cong   1 - \frac {\pi l |\omega|}{2R_c^2} \langle
|A(\varphi)|\rangle \cong \exp\left( - \frac {\pi l |\omega|}{2R_c^2} \langle
|A(\varphi)|\rangle \right).
\end{equation}
Finally, Eq.\ (\ref{fDFact}) yields
\begin{eqnarray}\label{fDDres}
&&
 f_{\tilde D'} \cong \frac 1 \pi \frac {\tilde D'_c}{ (\tilde D')^2 + (\tilde D'_c)^2 },
 \nonumber\\
&& \tilde D'_c \equiv \frac {\pi l }{2R_c^2}\langle |A(\varphi)|\rangle = \frac{\pi^2 c l}2
\langle |A(\varphi)|\rangle.
\end{eqnarray}
In contrast to the random-dislocation model which is characterized by Gaussian fluctuations of
the uniaxial anisotropy $D$, here the distribution of $D$ is Lorentzian and its reduced width
$\tilde D'_c$ is by a factor $l/R_c \ll 1$ smaller than that of Eq.\ (\ref{fDoneRes}).
The asymptote $f_{\tilde D'} \cong (1/\pi)\tilde D'_c/(\tilde D')^2$ at $\tilde D' \gg \tilde
D'_c$ is due to a single dislocation pair, as can be checked by an independent calculation
similar to that for random dislocations [cf.\ Eq.\ (\ref{fDoneRes})].
For edge dislocations running perpendicular to the $z$-axis and $g_0=g'_0$, one
has $\langle |A(\varphi)|\rangle = 3^{3/2}g_0/[4\pi^2(1-\sigma)]$, which for
$g_0=1$ and $\sigma=0.25$ yields $\langle |A(\varphi)|\rangle \approx 0.175$.
For the concentration of dislocations $c=10^{-3}$, which is rather common, the average
distance between dislocations is according to Eq.\ (\ref{RcDef}) $R_c\approx 17.8$, in lattice
units.
For the size of the dislocation pair $l=5$ which satisfies the applicability condition $l
\lesssim R_c$ of Eq.\ (\ref{fDDres}) one obtains the reduced width of the distribution of the
anisotropy constant $\tilde D'_c \approx 4.3 \times 10^{-3}$ which is expectedly smaller than
that for randomly distributed single dislocations.

If the distance between dislocations in a dislocation pair $l$ is comparable
with the average distance between dislocations $R_c$, the distribution of $D$
will be neither Gaussian nor Lorentzian.
It still can be obtained numerically by the method described above.
A more realistic model should include distribution of the dislocation-pair
length $l$.
We do not attempt to consider these more complicated models here since too
little is known about the dislocations and their interaction in Mn$_{12}$ and
other molecular magnets.
More detailed experimental investigation of the tunneling resonance and EPR
lineshapes, as well as X-ray scattering investigations, are needed to elucidate
the distribution of dislocations in these materials.
Still, the first results reported on in this Letter show that dislocations at
reasonable concentrations can be made responsible for the experimentally
observed linewidths in Mn$_{12}$.


This work has been supported by the NSF Grant No.\ 9978882.

\end{document}